\documentstyle[bezier,emlines2,11pt]{article}

\title{\bf\hspace{6 cm}SPbU-IP-00-06\\
\vspace{1 cm}
Nucleus-nucleus scattering in
perturbative QCD with $N_c\rightarrow\infty$}
\author{M.Braun \\ Department of High Energy physics,
 University of S.Petersburg,\\
198904 S.Petersburg, Russia}
\def\beq{\begin{equation}}
\def\eeq{\end{equation}}
\def\noi{\noindent}

\begin{document}

\maketitle

\noi{\bf Abstract.}
In the perturbative QCD with $N_c\rightarrow\infty$
equations for the amplitude of the nucleus-nucleus scattering
are derived by the effective field method. The asymptotic form of the
solution is discussed. It is argued that in the high-energy limit
the total nucleus-nucleus cross-sections become constant and purely
geometrical.


\section{Introduction}
In the framework of the colour dipole model of A.H.Mueller [1,2]
it follows that in the high-colour limit $N_c\rightarrow\infty$ the scattering 
on a heavy nucleus is exactly described by the sum of fan diagrams constructed 
of BFKL pomerons, each of them splitting into two [3].
The equation for the sum of  BFKL fan diagrams was  first
written by I.Balitsky [4] in his original operator expansion formalism.
Then it was rederived by Yu.Kovchegov [5] in the colour dipole framework
and by the author by directly summing the BFKL fan diagrams [6].
The perturbative solution of this equation in the region of small
non-linearity (outside
the saturation region) was studied in [7]. Asymptotic estimates of the
solution were presented in [8]. Finally in [6] the exact solution of the
equation was obtained by direct numerical methods.
The main physical results following from these studies are that, first,
at high rapidities $Y$
the hA total cross-section saturates  to its geometrical
limit $2\pi R_A^2$ and, second, the gluon density in the nucleus
aquires a form of a soliton in $Y-\ln k$ space moving towards higher
momenta with nearly a constant velocity as  $Y$ increases. This last
property supports the applicability of the perturbative treatment,
since the well-know diffusion of the BFFKL pomeron towards small momenta
results to be stopped.

In the present paper we attempt to generalize these results to
nucleus-nucleus (AB) scattering. In this case, in
the $N_c\rightarrow\infty$
limit the total amplitude is given by the sum of all tree diagrams
constructed of BFKL pomerons  and the triple pomeron vertex.
In contrast to the hA case, the vertex now describes not only splitting of
a pomeron into two but also fusion of two pomerons into one. The diagrams
for the amplitude accordingly result much more complicated than the fan
diagrams relevant for the hA case. However, using the effective field
theory methods developed for summing such diagrams long ago [9,10], one can
construct an equation which describes the AB amplitude. Naturally this
equation  (in fact a pair of equations) results much more complicated
than for
the case of hA scattering. Its exact (numerical) solution does not
seem realistic.  However simple asymptotic estimates, analogous to the
ones made in [7,8 ], show that the total AB cross-section tends  to its
geometrical limit at high rapidities similar to the hA case.

Unfortunately the gluon density in the overlapping area cannot be
found from these estimates, but rather requires knowledge of the
solution in more detail. We leave this problem for future studies.

\section{Effective field theory for AB scattering}
At fixed overall impact parameter $b$ the AB 
amplitude ${\cal A}(Y,b)$ can be presented as an
exponential of its connected part:
\beq
{\cal A}(Y,b)=2is\left(1-e^{- T(Y,b)}\right)
\eeq
The dimensionless $T$ is an
integral over
two impact parameters $b_A$ and $b_B$ of the collision point relative to
the centers of the nuclei A and B:
\beq
T(Y,b)=\int d^2b_Ad^2b_B\delta^2(b-b_A+b_B)T(Y, b_A,b_B)
\eeq

As mentioned in the Introduction, in the perturbative QCD with
$N_c\rightarrow\infty$ the amplitude
$-T(Y,b_A,b_B)$ is given by a sum of all connected tree diagrams constructed
of BFKL pomerons and the triple pomeron vertex. More concretely, in these
diagrams a line ("propagator") connecting two points
$y_1,r_1$ and $y_2,r_2$ corresponds to one half of the forward
BFKL Green function [11]:
\beq
G(y_1-y_2,r_1,r_2)=\frac{r_1r_2}{32\pi^2}\theta(y_1-y_2)
\sum_{n=-\infty}^{+\infty}e^{in(\phi_1-\phi_2)}
\int\frac{d\nu e^{(y_1-y_2)\omega(\nu)}(r_1/r_2)^{-2i\nu}}
{\left(\nu^2+\frac{(n-1)^2}{4}\right)
\left(\nu^2+\frac{(n+1)^2}{4}\right)},
\eeq
where $\phi_{1,2}$ and are the azimuthal angles and
\beq
\omega(\nu)=\frac{\alpha_sN_c}{\pi}(\psi(1)-{\rm Re}
\psi(1/2+i\nu))
\eeq
are the BFKL levels.
Due to  the azimuthal symmetry of the projectile and target colour
densities one may retain only the term with zero orbital momenta
$n=0$ in (3). 

The interaction between the pomerons is realized via the triple pomeron
vertex. It is non-local and not symmetric in the incoming and outgoing
pomerons. Its form for the splitting of a pomeron into two was established
in [2,12,13]. At $N_c\rightarrow\infty$ for the transition $1\rightarrow 2+3$
the three BFKL Green functions are connected by it as  follows
(see Fig. 1a)
\beq
\frac{4\alpha_s^2N_c}{\pi}\int\frac{d^2r_1d^2r_2d^2r_3}{r_1^2r_2^2r_3^2}
\delta^2(r_1+r_2+r_3)G(y'_1-y, r'_1,r_1)\nabla_1^4r_1^4
G(y-y'_2, r_2,r'_2)G(y-y'_3, r_3,r'_3)
\eeq
Here it is assumed that the operator $\nabla_1$ acts on the left.
The form of the vertex for the fusion of two pomerons into one is
actually not known. However, the symmetry between target and projectile
prompts us to assume that for the inverse process $2+3\rightarrow 1$
the BFKL functions are to be joined as    (Fig. 1b)
\beq
\frac{4\alpha_s^2N_c}{\pi}\int\frac{d^2r_1d^2r_2d^2r_3}{r_1^2r_2^2r_3^2}
\delta^2(r_1+r_2+r_3)
G(y'_2-y, r'_2,r_2)G(y'_3-y, r'_3,r_3)
\nabla_1^4r_1^4G(y-y_1, r_1,r'_1)
\eeq

Finally we have to describe the interaction of the pomerons with the
two nuclei. The BFKL Green functions corresponding to the external
legs of the diagrams are to be integrated with the colour density of
each nucleus. We take the target nucleus at rest, that is, at rapidity
zero. Then each outgoing external BFKL Green function is to be transformed
into
\beq
g^2AT_A(b_A)\int d^2r' G(y,r,r')\rho_N(r')\equiv
\int dy'd^2r'G(y-y',r,r')\tau_A(y',r')
\eeq
where $\rho_N$ is the colour density of the nucleon, $T_A$ is the
profile function of the nucleus A and we define
\beq
\tau_A(y,r)=g^2AT_A(b_A)\rho_N(r)\delta(y)
\eeq
(with dependence on $b_A$ implicit). Similarly each ingoing BFKL external
Green function is transformed into
\beq
\int dy'd^2r'G(y'-y,r',r)\tau_B(y',r')
\eeq
where
\beq
\tau_B(y,r)=g^2BT_B(b_B)\rho_N(r)\delta(y-Y)
\eeq

To find the amplitude one has to sum over all connected
diagrams with $M$ ingoing and
$N$ outgoing lines, corresponding to $M$ interactions with the projectile
and $N$ interactions with the target, divided by $M!N!$.

It is trivial to see that this sum exactly corresponds to the sum of
tree diagrams generated by an effective quantum theory of two pomeronic
fields $\Phi(y,r)$ and $\Phi^{\dagger}(y,r)$ with 
action
\beq
S=S_0+S_I+S_E
\eeq
consisting of three terms, which correspond to free pomerons,
their mutual interaction and their
interaction with external sourses (nuclei) respectively.

To give the correct propagators $S_0$ has to be chosen as
\beq
S_0=2\int dy_1d^2r_1dy_2d^2r_2\Phi(y_1,r_1)G^{-1}
(y_1-y_2,r_1,r_2)\Phi^{\dagger}(y_2,r_2)\equiv
2\langle\Phi|G^{-1}|\Phi^{\dagger}\rangle
\eeq
where $\langle|\rangle$ means the integration over $y,r$.
Note that the sign of $S_0$ corresponds to the substitution
of the conventionally defined field variables $\Phi$ and $\Phi^{\dagger}$:
\beq
\Phi\rightarrow i\Phi,\ \ \Phi^{\dagger}\rightarrow i\Phi^{\dagger}
\eeq
which allows to make all terms of the action real.

According to (5), (6) the interaction term $S_I$ is local in rapidity
\beq
S_I=
\frac{4\alpha_s^2N_c}{\pi}\int dy
\int\frac{d^2r_1d^2r_2d^2r_3}{r_1^2r_2^2r_3^2}
\delta^2(r_1+r_2+r_3)
\Big[\nabla_1^4r_1^4 \Phi^{\dagger}(y,r_1)\Phi(y,r_2)\Phi(y, r_3)
\ \ +\ \ c.c.\ \Big]
\eeq
The overall sign combines the initial factor $i$ and $i^3$ from the
substitution (13).

Finally the interaction with the nuclei is local both in rapidity and
coordinates:
\beq
S_E=-\int dyd^2r\Big[\Phi(y,r)\tau_A(y,r)+\Phi^{\dagger}(y,r)\tau_B(y,r)
\Big]
\eeq
The minus sign comes from the initial $i$ and the substitution (13).

The amplitude $T(Y,b_A,b_B)$ is then expressed through a functional
integral
\beq
Z=\int D\Phi D\Phi^{\dagger}e^{S/\mu^2}
\eeq
where $\mu$ is an arbitrary mass scale necessary to adjust the dimensions
of various parts of the action. Keeping only the connected diagrams
one finds
\beq
T(Y,b_A,b_b)=-\mu^2\ln \frac{Z}{Z_0}
\eeq
where $Z_0$ is the value of $Z$ for $S_E=0$.
Functional integral $Z$ is to be calculated in the classical
approximation to retain only the tree diagrams. 
This  gives
\beq
T(Y,b_A,b_B)=-S_E\{\Phi,\Phi^{\dagger}\}=
\int d^2r\Big[\Phi(0,r)\hat{\tau}_A(r)+\Phi^{\dagger}(Y,r)\hat{\tau}_B(r)
\Big]
\eeq
where $\Phi$ and $\Phi^{\dagger}$ are the solutions of the classical
equation of motion and $\hat{\tau}$'s are (8) and (10) with
the $\delta$ functions of rapidity dropped. We see that the result is
independent  of the scale $\mu$, as it should be.

\section{Equations of motion}
Before writing out the classical equation of motion, we transform
the functional integral (16) to new variables in which the non-locality
of the Lagrangian becomes substantially reduced.
We put
\beq
\Phi(y,r)=r^2\phi(y,r),\ \ \Phi^{\dagger}(y,r)=r^2\phi^{\dagger}(r,y)
\eeq
In these variables the interaction term becomes
\beq
S_I=
\frac{4\alpha_s^2N_c}{\pi}\int dy
\int d^2r_1d^2r_2d^2r_3
\delta^2(r_1+r_2+r_3)
\Big[ K_1\phi^{\dagger}(y,r_1)\phi(y,r_2)\phi(y, r_3)
\ \ +\ \ c.c.\ \Big]
\eeq
where (dimensionles) operator $K$ has the form
\beq
K=r^2\nabla^4_rr^2
\eeq
Now we transform (20) to the momentum space to obtain
\beq
S_I=
\frac{4\alpha_s^2N_c}{\pi}\int dy
\int\frac{d^2q}{(2\pi)^2}
\Big[ K\phi^{\dagger}(y,q)\phi(y,q)\phi(y,q)
\ \ +\ \ c.c.\ \Big]
\eeq
with operator $K$ local in the momentum space
\beq
K=\nabla_q^2q^4\nabla_q^2
\eeq
As we observe, the interaction has become local in the momentum
space.

Now we turn to the free part $S_0$. In new variables it takes the form
\beq
S_0=
2\langle\phi|r^2G^{-1}r^2|\phi^{\dagger}\rangle
\eeq
It was shown in [6] that
\beq
r_1^2\nabla_1^4G(y,r_1,r_2)=g(y,r_1,r_2)r_2^2
\eeq
where $g$ is the Green function of the BFKL equation for the so-called
semi-amputated wave function:
\beq
\left(\frac{\partial}{\partial y}+H_1\right)g(y,r_1,r_2)=
\delta(y)\delta^2(r_1-r_2)
\eeq
Here $H_1$ is the BFKL Hamiltonian [11] acting on $r_1$.
We rewrite (26) in the operatorial form
\beq
Kr^{-2}Gr^{-2}=\left(\frac{\partial}{\partial y}+H\right)^{-1}
\eeq
wherefrom
\beq
r^{2}G^{-1}r^{2}K^{-1}=\frac{\partial}{\partial y}+H
\eeq
and finally
\beq
r^{2}G^{-1}r^{2}=\left(\frac{\partial}{\partial y}+H\right) K
\eeq
Since both $G(y,r_1,r_2)$ and $g(y,r_1,r_2)$ are symmetric in $r_1, r_2$
we also find in the same manner
\beq
r^{2}G^{-1}r^{2}=K\left(\frac{\partial}{\partial y}+H\right) 
\eeq
so that $K$ commutes with $H$. Using (29) and (30) in (24) we see that the
free part has become local in rapidity and can be expressed via the
BFKL Hamiltonian $H$:
\beq
S_0=2\langle\phi|K\left(\frac{\partial}
{\partial y}+H\right) |\phi^{\dagger}\rangle
\eeq
This part remains non-local both in the coordinate and momentum spaces
due to the non-locality of $H$.

The interaction with the nucleus part in the new variables takes the form
\beq
S_E=-\langle w_A|\phi\rangle-\langle\phi^{\dagger}|w_B\rangle
\eeq
where, in the coordinate space,
\beq
w_{A,B}(y,r)=r^2\tau_{A,B}(y,r)
\eeq

Now, with the action nearly completely local (except for the Hamiltonian
term) we can write out the equation of motion. We find
\beq
\frac{\delta S}{\delta\phi(y,q)}=
2K\left(\frac{\partial}{\partial y}+H\right)\phi^{\dagger}(y,q)
+\frac{4\alpha_s^2N_c}{\pi}
\Big( K{\phi^{\dagger}}^2(y,q)+2\phi(y,q)K\phi^{\dagger}(y.q)\Big)
-w_A(y,q)=0
\eeq
and
\beq
\frac{\delta S}{\delta\phi^{\dagger}(y,q)}=
2K\left(-\frac{\partial}{\partial y}+H\right)\phi (y,q)
+\frac{4\alpha_s^2N_c}{\pi}
\Big( K\phi^2(y,q)+2\phi^{\dagger}(y,q)K\phi(y.q)\Big)
-w_B(y,q)=0
\eeq
Applying operator $(1/2)K^{-1}$ from the left we find our final equations
of motion
\beq
\left(\frac{\partial}{\partial y}+H\right)\phi^{\dagger}(y,q)
+\frac{2\alpha_s^2N_c}{\pi}
\Big( {\phi^{\dagger}}^2(y,q)+2K^{-1}[\phi(y,q)K\phi^{\dagger}(y.q)]\Big)
-\frac{1}{2}K^{-1}w_A(y,q)=0
\eeq
and
\beq
\left(-\frac{\partial}{\partial y}+H\right)\phi(y,q)
+\frac{2\alpha_s^2N_c}{\pi}
\Big( \phi^2(y,q)+2K^{-1}[\phi^{\dagger}(y,q)K\phi(y.q)]\Big)
-\frac{1}{2}K^{-1}w_B(y,q)=0
\eeq

As we see the resulting equations are rather complicated, since they
involve nonlocal terms, bilinear in $\phi$ and $\phi^{\dagger}$,
which interconnect the two equations. Summing fan diagrams in the
hA case in fact leads to the the same equations, in which
however $w_B=0$. Then one immediately finds that $\phi=0$ identically,
which converts the first equation into
\beq
\left(\frac{\partial}{\partial y}+H\right)\phi^{\dagger}(y,q)
+\frac{2\alpha_s^2N_c}{\pi}{\phi^{\dagger}}^2(y,q)
-\frac{1}{2}K^{-1}w_A(y,q)=0
\eeq
This local equation is just the one studied in [4-6].

To conclude this section we present the non-local terms in Eqs. (36)
and (37) in a more explicit way. To this end we calculate the kernel of the
operator $K^{-1}$ in the momentum space. We have in the coordinate space
\beq
K^{-1}=r^{-2}\nabla_r^{-4}r^{-2}
\eeq
Using the identity [6]
\beq
\nabla_1^4G(0,r_1,r_2)=\delta^2(r_1-r_2)
\eeq
we can write the kernel  of $K^{-1}$ in the coordinate space as
\beq
K^{-1}(r_1,r_2)=r_1^{-2}G(0,r_1,r_2)r_2^{-2}
\eeq
Fourier transforming (41) and using (3) we find the kernel in the
momentum space:
\beq
K^{-1}(q_1,q_2)=\int \frac{d^2r_1d^2r_2}{r_1^2r_2^2}
e^{iq_2r_2-iq_1r_1}G(0,r_1,r_2)=
\frac{1}{8}\int \frac{d\nu}{(\nu^2+1/4)^2}I(\nu,q_1)I^{\ast}(\nu,q_2)
\eeq
where
\beq
I(\nu,q)=\int_0^{\infty} drr^{-2i\nu}J_0(qr)=
2^{-2i\nu}q^{-1+2i\nu}\frac{\Gamma(1/2-i\nu)}{\Gamma(1/2+i\nu)}
\eeq
Doing the integral over $\nu$ we finally find
\beq
K^{-1}(q_1,q_2)=\frac{\pi}{2}\frac{1}{q_>^2}
\left(\ln\frac{q_>}{q_<}+1\right)
\eeq
where $q_{>(<)}=\max (\min)\{q_1,q_2\}$.

Using (44) we can rewrite the nonlocal term in Eq. (36) as
\beq
4\alpha_s^2N_c
\int\frac{d^2q_1}{(2\pi)^2q_>^2}
\left(\ln\frac{q_>}{q_<}+1\right)
\phi(y,q_1)\nabla_1^2q_1^4\nabla_1^2\phi^{\dagger}(y,q_1)
\eeq
where $q_{>(<)}=\max (\min)\{q,q_1\}$.
The nonlocal term in Eq. (37) is obtained by complex conjugation.

\section{The total cross-section}
The obtained equations which determine the classical fields $\phi$ and
$\phi^{\dagger}$ are very difficult to solve even numerically.
The trouble lies not so in the their explicit non-locality, but in the
appearence of two different sources at two different rapidities.
Due to conditions $\phi=0$ for $y>Y$ and $\phi^{\dagger}=0$ for $y<0$
and the $\delta$-like dependence of the sources on rapidities,
one can drop the sources in Eq. (36) and (37) substituting them
by conditions
\beq
\phi^{\dagger}(y,q)_{y=0}=K^{-1}\hat{w}_A(q),\ \ 
\phi(y,q)_{y=Y}=K^{-1}\hat{w}_B(q)
\eeq
In the hA case one has only the first of these conditions, which
converts Eq. (36) into an evolution equation in rapidity, relatively
easily solved by conventional methods. As mentioned the non-local term is
zero in this case, but its presence would only slightly complicate the
solution. After all the BFKL Hamiltonian is also non-local (although
linear).

For the nucleus-nucleus scattering we have to solve homogeneous
Eqs (36) and (37) with both conditions (46) imposed upon the solution.
The Cauchy problem is thus transformed into an essentially more
difficult boundary  problem. A possible method of the solution is to
transform Eqs. (36) and (37) into a system of two non-linear
integral equations
in combined rapidity-momentum space, which one may try to solve
by iterations. Our experience in the hA problem
shows that for the solution to have a reasonable precision one requires
at least 800 points in the momentum and 400 points for 5  units of rapidity.
Thus to find  the amplitude for say $Y=15$ one has to perform
$1200\times 800^3\sim 10^{10}$ operations per iteration. On top of that the
convergence properties of the iteration procedure is unknown.

Here we shall not attempt at solving Eqs. (36) and (37) with any reasonable
degree of precision at all values of rapidity and momentum. Instead we
shall again use our experience with the case of hA scattering (fan diagrams),
where at any fixed momentum and $y\rightarrow\infty$ the solution
$\phi^{\dagger}(y,q)$ aquires a simple form, independent of the target
properties
\beq
\phi^{\dagger}(y,q)_{y\rightarrow\infty}=\frac{2\pi}{g^2}\ln\frac{Q(y)}{q}
\eeq
with $\ln Q(y)\simeq 2.34(\alpha_sN_c/\pi)y$. Note that (47) is not the
solution at all $y$ and $q$. In particular (47) is not valid at $q\sim Q$,
which is just the region which determines the gluon density. However
(47) is sufficient to establish that  the  hA scattering cross-section
tends to its geometric limit at high $y$ [7].

Our guess is that also in the nucleus-nucleus case function
$\phi^{\dagger}(y,q)$ aquires the form (47) at large
rapidities $y\sim Y$ and $\phi(y,q)$ aquires the same form with
$y\rightarrow Y-y$.
To support this behaviour we are going to demonstrate that in these
limits the non-local terms in Eqs. (36) and (37) can be neglected, so that
the equations decouple and become similar to the hA case.

Actually the demonstration is quite simple. Take Eq. (36) at large $y$
and put the conjectured asymptotics (47) into it. According to (45)
the mixing non-local term will then be given by
\beq
\alpha_sN_c
\int\frac{d^2q_1}{(2\pi)^2q_>^2}
\left(\ln\frac{q_>}{q_<}+1\right)
\phi(y,q_1)\nabla_1^2q_1^4\nabla_1^2\ln\frac{Q(y)}{q_1}
\eeq
Note that $\phi(y,q_1)$ enters at large $y$ and small $Y-y$. Its exact form is
unknown but we can safely assume that it rapidly falls with $q_1$ similar
to the inhomogeneous term $K^{-1}w_B$ from which it is separated by
a relatively small distance in rapidity.
Action of the operator $K=\nabla^2q^4\nabla^2$ on the
asymptotic form of $\phi^{\dagger}$ however gives zero. In fact
\[ \nabla_1^2\ln\frac{Q(y)}{q_1}=-2\pi\delta^2(q_1)\]
Subsequent integration over $q_1$ gives zero at any finite $q$ due to factor
$q_1^4$.
Thus the non-local term  is zero in Eq. (36) at $y\sim Y\rightarrow\infty$.
Therefore the equation aquires the same form as for hA scattering (which
implies that fan diagrams going from top to bottom dominate). This
means that the asymptotical behaviour (47) is indeed true.
The same result is found for the non-local term in Eq. (37) at small $y$ and
$Y-y\rightarrow\infty$ assuming the asymptotic form (47) for $\phi(Y-y,q)$.
Its meaning is that  fan diagrams going from bottom to top
dominate in this limit.

Functions $\Phi(y,r)$ and $\Phi^{\dagger}(y,r)$ which
actually determine the amplitude according to (18) are related to
$\phi(y,q)$ and $\phi^{\dagger}(y,q)$ by
\beq
\Phi(y,r)=-\int \frac{d^2q}{(2\pi)^2}e^{iqr}\nabla_q^2 \phi(y,q)
\eeq
and similarly for the conjugated function. Using the asymptotical
expression (47) we then get at high $Y$
\beq
\Phi(0,r)=\frac{1}{g^2}\theta(R_B-b_B),\ \ \Phi^{\dagger}(Y,r)=
\frac{1}{g^2}\theta(R_A-b_A)
\eeq
The two $\theta$ functions appear because according to
Eqs. (36) and (37) $\phi=0$ ($\phi^{\dagger}=0$) when $w_B=0$ ($w_B=0$),
that is for $b_B>R_B$ ($b_A>R_A$).

Putting (50) in (18) we obtain the connected part of the amplitude
at $Y>>1$ as
\beq
T(Y,b_A,b_B)=AT_A(b_A)\theta(R_B-b_B)+
T_B(b_B)\theta(R_A-b_A)
\eeq
It results independent of $Y$.
After the integration over $b_B$ and $b_B$  we find
\beq
T(Y,b)=\int d^2b_Ad^2b_B\delta^2(b-b_A+b_B)
\theta(R_A-b_A)\theta(R_B-b_B)[AT_A(b_A)+BT_B(b_B)]
\eeq
According to (1) the total AB cross-section is given by
\beq
\sigma^{tot}(Y)=2\int d^2b \left(1- e^{-T(Y,b)}\right)
\eeq
>From (52) and (53) one concludes that at large $Y$ the cross-section
does not depend on $Y$. 
It saturates at a value which is purely geometrical and
for $A>>1$ or/and $B>>1$.
approaches the black disk limit in the overlap area.

\section{Conclusions}
We have derived a pair of equations which describe the nucleus-nucleus
scattering in the perturbative QCD with a large number of colours (or,
alternatively in the quasi-classical limit, or, in the limit
$A,B\rightarrow\infty$). The equations contain mixing terms which are
both non-linear and non-local. In contrast to the hA case the
equations are
to be solved with given boundary conditions at rapidities both of
the projectile and target, which complicates the solution enormously.

However the asymptotical form of the solution at fixed momentum and
large rapidities is shown  to be the same as for the hA case.
This allows to demonstrate that at large rapidities the total AB
cross-section becomes independent of energy and given by purely
geometric considerations. At large $A$ or/and $B$ it corresponds to
the scattering on  the black disc in the overlap area.

Going to particle production in AB collisions, the situation at central
rapidities seems to be rather simple. The inclusive cross-section will
be described by diagrams like shown in Fig. 2, with the target and
projectile parts joined by a single pomeron, from which the observed
particle is emitted. Evidently this contribution is just a convolution
of the production vertex with two gluon densities of the projectile and
target, each one corresponding  to fan diagrams and found in [6].
At rapiditiy distances from the target or projectile
$\delta y\sim 1/\Delta$ where $\Delta$
is the pomeron intercept  the problem
does not look so simple, since the AGK rules are rather complicated
in this region (see e. g. [14]) and the usual cancellation of all diagrams
except of the structure shown in Fig. 2 is not at all obvious.
We leave this problem for future studies.

\section{References}
\noi
1. A.Mueller, Nucl. Phys.,{\bf B415} (1994) 373.\\
2. A.Mueller and B.Patel, Nucl. Phys.,{\bf B425} (1994) 471.\\
3. M.A.Braun and G.P.Vacca, Eur. Phys. J {\bf C6} (1999) 147.\\
4. I.Balitsky, hep-ph/9706411; Nucl. Phys. {\bf B463} (1996) 99.\\
5. Yu. Kovchegov, Phys. Rev {\bf D60} (1999) 034008.\\
6. M.Braun, preprint LU TP 00-06 (hep-ph/0001268)\\
7. Yu. Kovchegov, preprint CERN-TH/99-166 (hep-ph/9905214).\\
8 .E.Levin and K.Tuchin, preprint DESY 99-108, TAUP 2592-99
(hep-ph/9908317).\\
9. A.Schwimmer, Nucl. Phys. B94 (1975)445.\\
10. D.Amati, L.Caneshi and R.Jengo, Nucl. Phys. {\bf B101} (1975) 397.\\
11. L.N.Lipatov in: "Perturbative QCD", Ed. A.H.Mueller, World Sci.,
Singapore (1989) 411.\\
12. J.Bartels and M.Wuesthoff, Z.Phys., {\bf C66} (1995) 157.\\
13. M.A.Braun, Eur. Phys. J {\bf C6} (1999) 321.\\
14. M.Ciafaloni {\it et al.}, Nucl. Phys. {\bf B98} (1975) 493.\\

\section{Figure captions}
\noi Fig. 1. The triple pomeron vertex for the splitting of a
pomeron into two (a) and fusion of two pomerons into one (b).\\
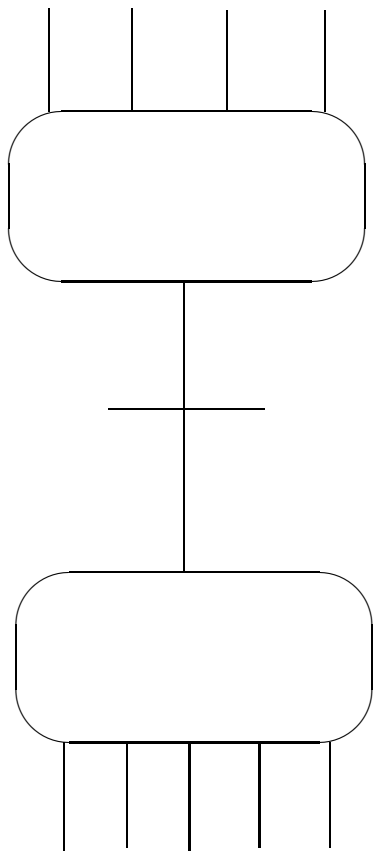
Fig. 2. The generic diagram for the inclusive particle production
in the central region in AB collisions\\
\newpage
\unitlength=1.00mm
\special{em:linewidth 0.4pt}
\linethickness{0.4pt}
\begin{picture}(117.34,136.33)
\put(13.00,115.83){\oval(20.67,14.33)[]}
\put(-8.00,77.83){\oval(20.67,14.33)[]}
\put(36.00,78.50){\oval(20.67,14.33)[]}
\put(64.67,115.83){\oval(20.67,14.33)[]}
\put(107.00,116.16){\oval(20.67,14.33)[]}
\put(85.67,78.50){\oval(20.67,14.33)[]}
\put(9.00,123.33){\line(0,1){9.67}}
\put(17.33,123.33){\line(0,1){9.33}}
\put(60.33,123.00){\line(0,1){9.67}}
\put(69.33,123.00){\line(0,1){9.67}}
\put(102.33,123.33){\line(0,1){9.33}}
\put(112.00,123.33){\line(0,1){9.00}}
\put(-12.67,71.00){\line(0,-1){11.00}}
\put(-3.33,70.66){\line(0,-1){10.33}}
\put(31.67,71.33){\line(0,-1){10.67}}
\put(41.00,71.33){\line(0,-1){10.67}}
\put(81.33,71.33){\line(0,-1){10.33}}
\put(90.33,71.33){\line(0,-1){10.67}}
\put(8.33,108.66){\line(0,-1){8.67}}
\put(18.00,108.66){\line(0,-1){8.33}}
\put(8.33,100.00){\line(-1,-1){14.67}}
\put(18.00,100.33){\line(1,-1){15.33}}
\put(12.67,92.00){\line(-1,-1){11.33}}
\put(12.67,92.00){\line(1,-1){13.33}}
\put(81.67,85.66){\line(0,1){9.33}}
\put(90.00,85.66){\line(0,1){9.00}}
\put(81.67,95.00){\line(-1,1){13.67}}
\put(90.00,94.33){\line(1,1){14.33}}
\put(85.33,103.33){\line(-1,1){10.67}}
\put(85.33,103.33){\line(1,1){11.33}}
\put(-1.00,135.66){\makebox(0,0)[lc]{$y'_1,r'_1$}}
\put(52.00,136.00){\makebox(0,0)[lc]{$y'_2,k'_2$}}
\put(98.33,136.33){\makebox(0,0)[lc]{$y'_3,k'_3$}}
\put(12.33,116.66){\makebox(0,0)[cc]{G}}
\put(64.00,116.00){\makebox(0,0)[cc]{G}}
\put(106.33,116.33){\makebox(0,0)[cc]{G}}
\put(-9.00,101.00){\makebox(0,0)[lc]{$y$}}
\put(64.33,96.33){\makebox(0,0)[lc]{$y$}}
\put(13.00,104.00){\makebox(0,0)[cc]{$r_1$}}
\put(85.67,90.33){\makebox(0,0)[cc]{$r_1$}}
\put(2.67,89.00){\makebox(0,0)[cc]{$r_2$}}
\put(22.67,88.33){\makebox(0,0)[cc]{$r_3$}}
\put(75.67,106.66){\makebox(0,0)[cc]{$r_2$}}
\put(95.00,106.66){\makebox(0,0)[cc]{$r_3$}}
\put(-9.00,78.33){\makebox(0,0)[cc]{G}}
\put(35.33,78.66){\makebox(0,0)[cc]{G}}
\put(85.67,78.66){\makebox(0,0)[cc]{G}}
\put(-8.33,56.00){\makebox(0,0)[cc]{$k'_2$}}
\put(36.00,55.66){\makebox(0,0)[cc]{$k'_3$}}
\put(85.67,55.33){\makebox(0,0)[cc]{$k'_1$}}
\put(13.00,41.00){\makebox(0,0)[cc]{a}}
\put(85.33,39.33){\makebox(0,0)[cc]{b}}
\put(48.67,23.33){\makebox(0,0)[cc]{{\Large Fig. 1}}}
\end{picture}
\newpage
\unitlength=1mm
\special{em:linewidth 0.4pt}
\linethickness{0.4pt}
\begin{picture}(94.00,142.33)
\put(69.33,117.33){\oval(47.33,22.67)[]}
\put(70.33,56.00){\oval(47.33,22.67)[]}
\put(69.00,106.00){\line(0,-1){38.67}}
\put(59.00,89.00){\line(1,0){20.67}}
\put(51.00,128.67){\line(0,1){13.67}}
\put(87.67,128.67){\line(0,1){13.33}}
\put(62.00,128.67){\line(0,0){0.00}}
\put(62.00,128.67){\line(0,1){13.67}}
\put(74.67,128.67){\line(0,1){13.33}}
\put(53.00,44.67){\line(0,-1){14.33}}
\put(88.33,44.67){\line(0,-1){14.00}}
\put(61.33,44.67){\line(0,-1){14.00}}
\put(79.00,44.67){\line(0,-1){14.00}}
\put(69.67,44.67){\line(0,-1){14.33}}
\put(69.67,8.33){\makebox(0,0)[cc]{{\Large Fig. 2}}}
\end{picture}
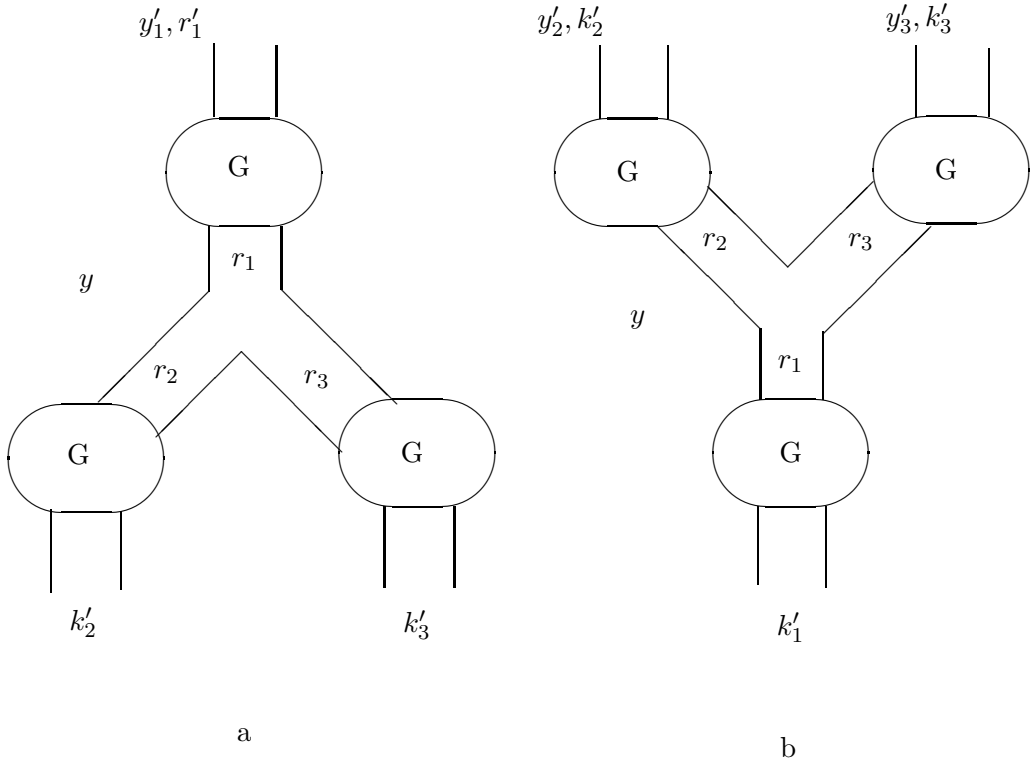
\end{document}